# Hardware and Logic Implementation of Multiple Alarm System for GSM BTS Rooms

Arifa Ferdousi[1] and Sadeque Reza Khan[2]

[1]Dept. of Computer Science and Engineering, Varendra University, Rajshahi, Bangladesh
arifaferdousi@yahoo.com

[2]Dept. of Electronic and Communication Engineering, NITK, India
sadeque_008@yahoo.com

## ABSTRACT

*Cellular communication becomes the major mode of communication in present century. With the development of this phase of communication the globalization process is also in its peak of speed. The development of cellular communication is largely depending on the improvement and stability of Base Transceiver Station (BTS) room. So for the purpose of the development of cellular communication a large numbered BTS rooms are installed throughout the world. To ensure proper support from BTS rooms there must be a security system to avoid any unnecessary vulnerability. Therefore multiple alarm system is designed to secure the BTS rooms from any undesired circumstances. This system is designed with a PIC Microcontroller as a main controller and a several sensors are interfaced with it to provide high temperature alarm, smoke alarm, door alarm and water alarm. All these alarms are interfaced with the alarm box in the BTS room which provides the current status directly to Network Management Centre (NMC) of a Global System for Mobile (GSM) communication network.*

## 1. INTRODUCTION

BTS rooms are generally sealed with sophisticated and obviously costly electrical and electronics devices which need constant inspection to ensure that those devices are safe and operational [1]. Studying the location of these rooms it can be found that these rooms are located in such a position where rain water, illegal human intrusion, fire and many other unexpected incidents can occur. So remote monitoring is a must needed service for these rooms. An intelligent alarm system can easily assist in this phenomenon by studying all the factors of keeping the room safe from the unexpected situations. The system should be such that it can monitor room temperature all the time and generates alarm when the temperature goes above the set temperature as the ambient temperature requirement is -5˚C ~ +45˚C and is recommended to be always kept between 15˚C ~ 30˚C [2]. It also can detect if any kind of smoke, hence a fire event occurs in the room and ultimately generates an alarm. System is also equipped with sensors which can detect any kind water seepage in the room. Any kind of human intrusion through the door can also be easily tracked by this system. So in this project an intelligent alarm system is developed with 8-bit 40 pin PIC microcontroller IC 16F877A and combination of some sensors which generates alarms for any unexpected intrusions and sends the alarms to the Network Management Centre (NMC) [3] which is under Operation and Support Subsystem (OSS) [4] through an alarm box positioned in BTS room.

## 2. PROJECTED SYSTEM

Proposed system is built with a main controller that is PIC microcontroller 16F877A and some combination of sensors. This alarm system is designed such that illegal Human Intrusion can be detected, water seepage be monitored, smoke/Fire alarm can be generated and room temperature can be monitored via different sensors.

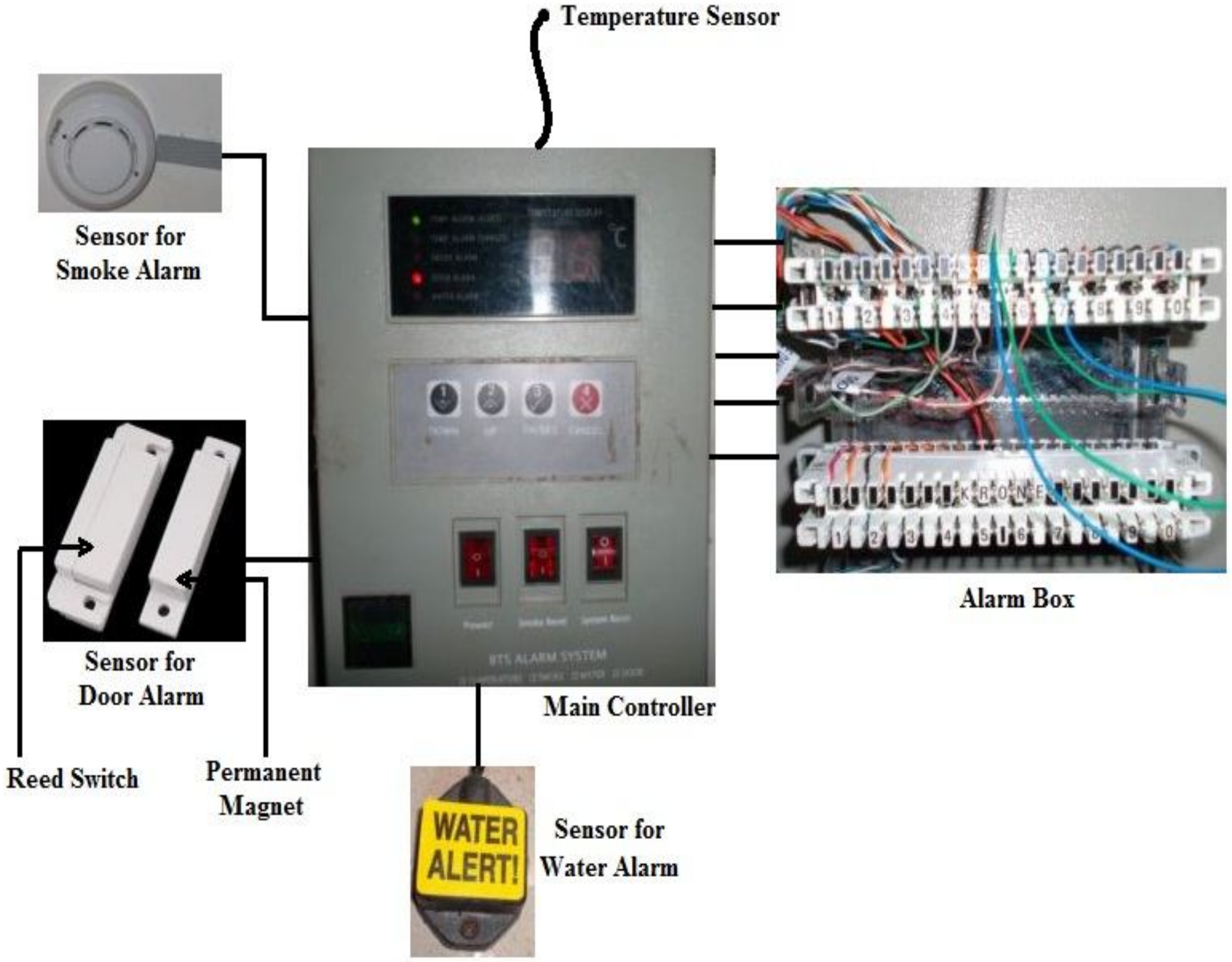

Figure 1. Proposed System.

## 3. HARDWARE DESIGN

### 3.1. Control Module

The control module is built with a microcontroller IC. The central controller is Microchip PIC16F877A. It consists of 40 I/O (Bi directional lines) with 25mA current in per pin and operating voltage range of 2.0V to 5.5V [5], [6]. It also has eight channel built-in A/D converter, serial communication and data EEPROM.

### 3.2. Temperature Sensor

The LM35 is an integrated circuit sensor that can be used to measure temperature with an electrical output proportional to the Celsius (Centigrade) temperature [7]. Temperature is directly measured by the AN0 pin of microcontroller IC. The value of temperature from the sensor is calculated from the given equation 1. If the temperature value is less than the value preset is EEPROM0 then it is assumed as normal temperature. Again the temperature is greater than the value of EEPROM0 then it is alert temperature and for the temperature that is greater than the value in EEPROM1 the status is danger temperature.

$$\text{Temperature} = (\text{ADC0 (AN0)} * 4/8)\ ^{o}\text{C} \quad \text{................................ Eqn.1}$$

If Temperature < Temperature (EEPROM0): Normal Temperature

If Temperature > Temperature (EEPROM0): Alert Temperature

If Temperature > Temperature (EEPROM2): Danger Temperature

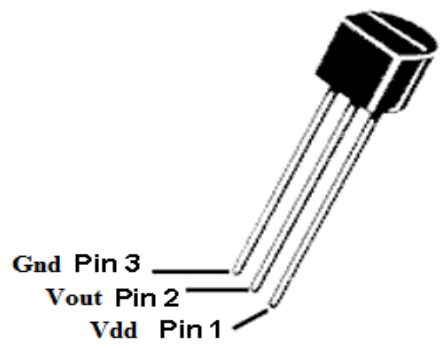

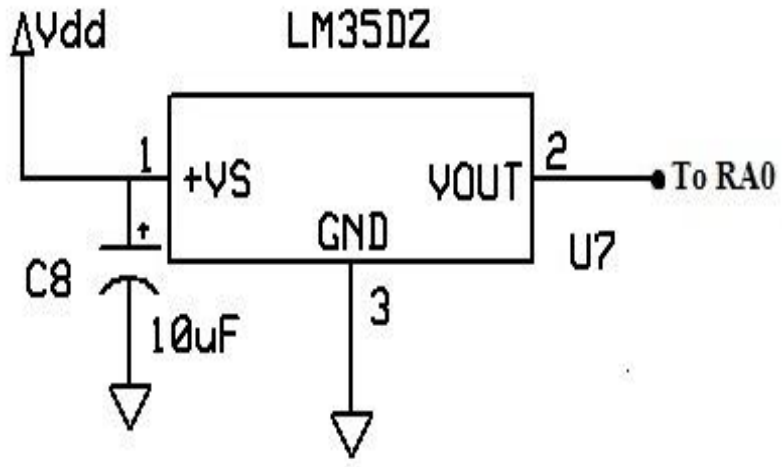

Figure 2. Pin Configuration and Connection diagram of LM35DZ

### 3.3. Water Alarm Section

Water alarm is necessary in a BTS room to keep it away from flood. As water is harmful for the sophisticated electronics components in BTS room so whenever water inserts in the floor of BTS room it is necessary to be informed to the authority to take essential steps. A water alarm section can be easily designed by using a transistor [8] shown in figure 3. Only two wires will come out from the controller to the floor of BTS room to give a sense of presence of water to the main controller. This section can also be designed with an Ultrasound Sensor [9] but the sensor is costly enough to implement in this controller.

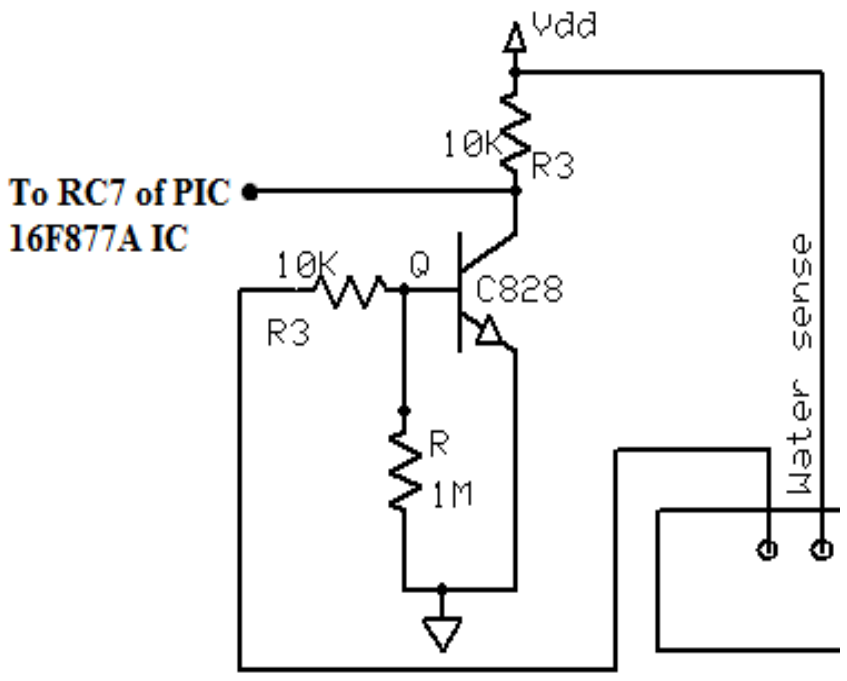

Figure 3. Water Alarm Circuit

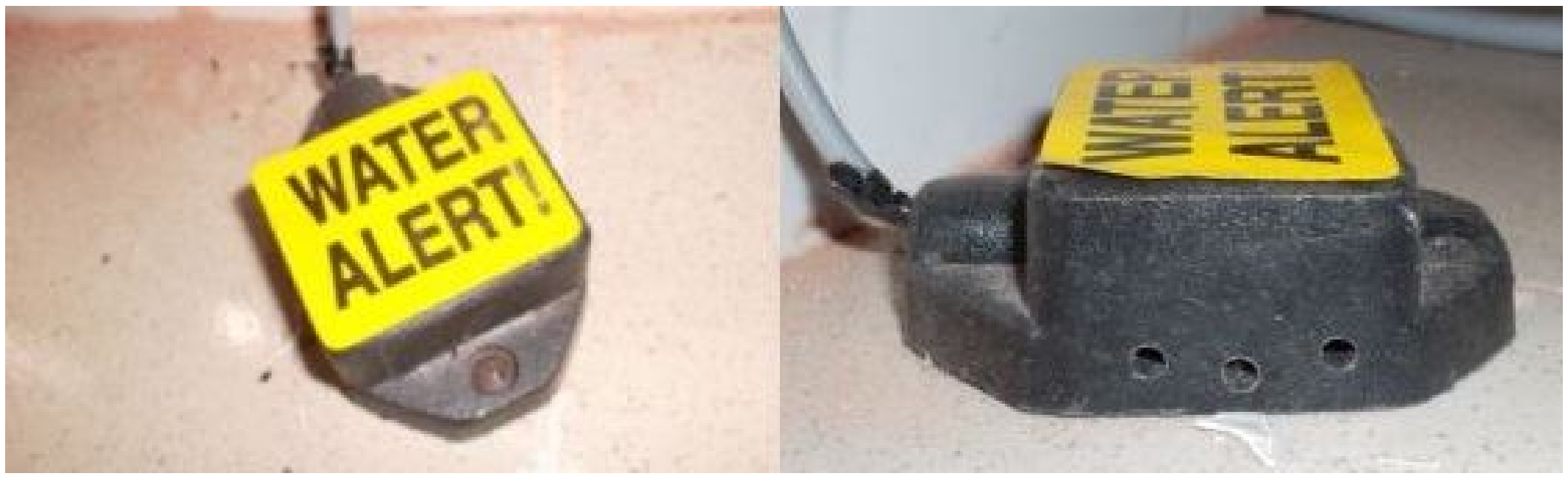

Figure 4. Placement of Wires Coming from Controller to Floor of BTS

### 3.4. Smoke/ Fire Alarm Section

Fire detection systems available currently are primarily smoke detectors [10]. The combination of the rates of rise of smoke and either carbon monoxide or carbon dioxide concentration provides a potential fire alarm algorithm to increase the reliability of smoke detectors, and to reduce the time to alarm. Photoelectric sensors are mostly used in smoke detectors as a precise means of sensing. Photoelectric sensor technology relies on an electric current that produces a beam of light [11]. When the beam of light is interrupted, an alarm sounds. Photoelectric sensors detect slow, smouldering, smoky fires more quickly than other technologies. These detectors contain a light source and a light-sensitive electric cell. Smoke particles in the detector deflect the light onto the photoelectric cell, thus generating a current and triggering the alarm. For this project a smoke detector of renowned company is used where the sensing is taken directly from the sensor and interfaced into the controller.

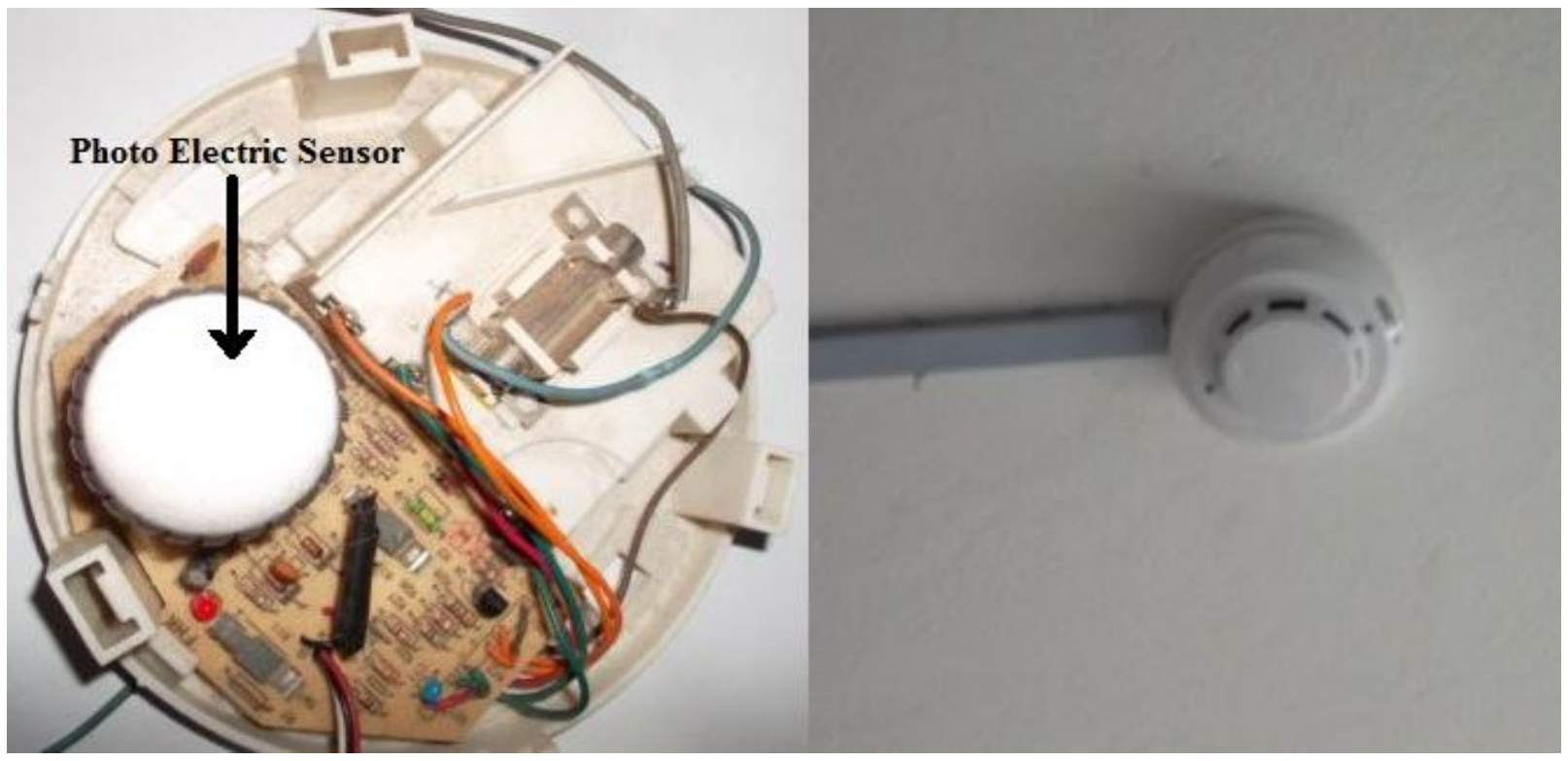

Figure 5. Internal structure of smoke alarm device and wall mounted sensor device

### 3.5. Door Alarm Section

Door alarm is developed with a Reed switch [12] and a permanent magnet. Permanent magnet is installed in the moving part of the door and Reed switch is in the fixed part of the door. So when ever door is opened a sense will go to the controller to generate alarm. The sense is taken by the controller through an opto-coupler.

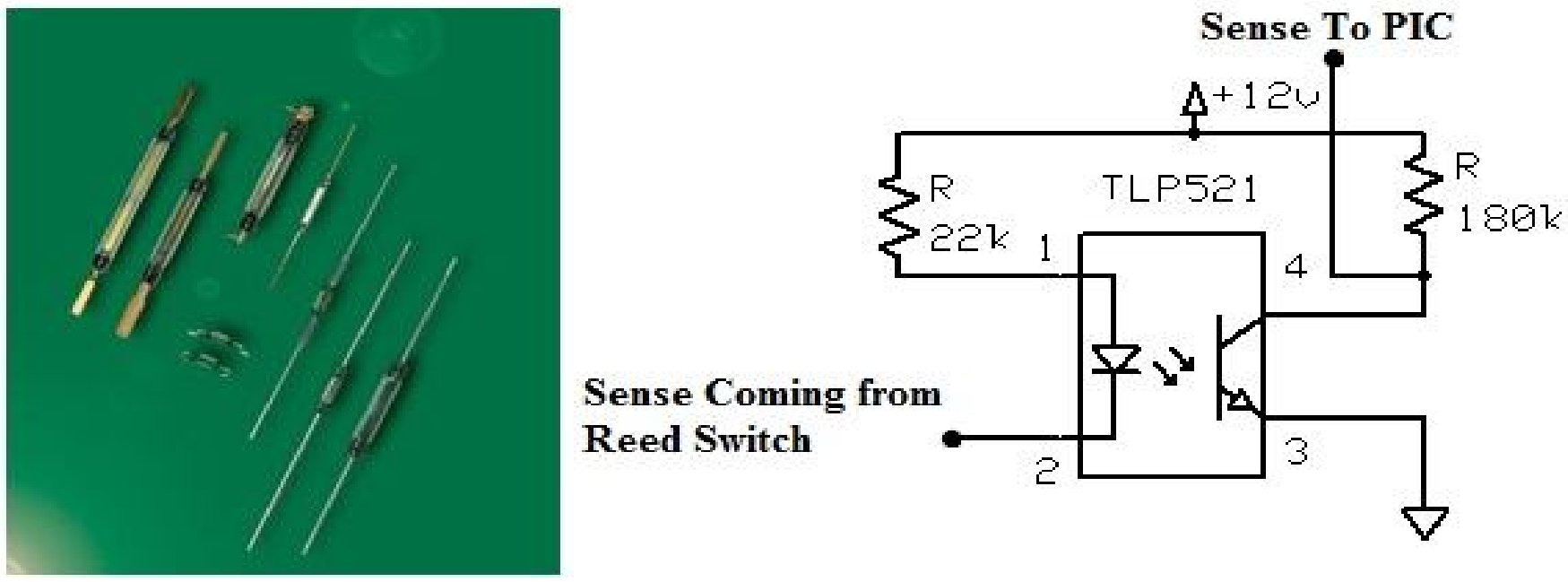

Figure 6. Reed Switch and circuit for taking the sense from reed switch to PIC

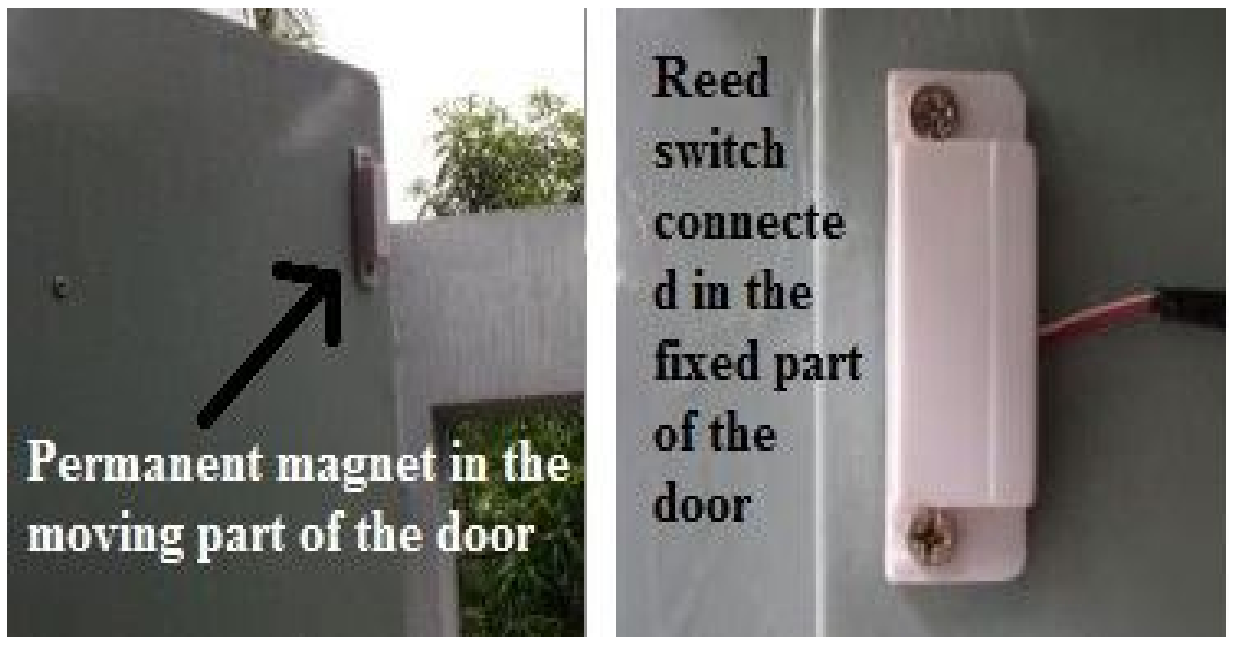

Figure 7. Implementation of door sensor

## 4. Logic Development

“Flowcode Ver. 5”, advanced simulation software is used to develop this project. It is a kind of software which has made the programming easy and more accurate rather than other old PICmicro chip software’s by using the highest level of programming [13], [14]. For this project the component macros Analog to Digital Converter, Seven Segment Display, Switch, LED and EEPROM are used in the following software which is shown in figure 8. The software Flowcode Ver. 5 provides a real time simulation which helps a lot in developing this project before going to practical hardware.

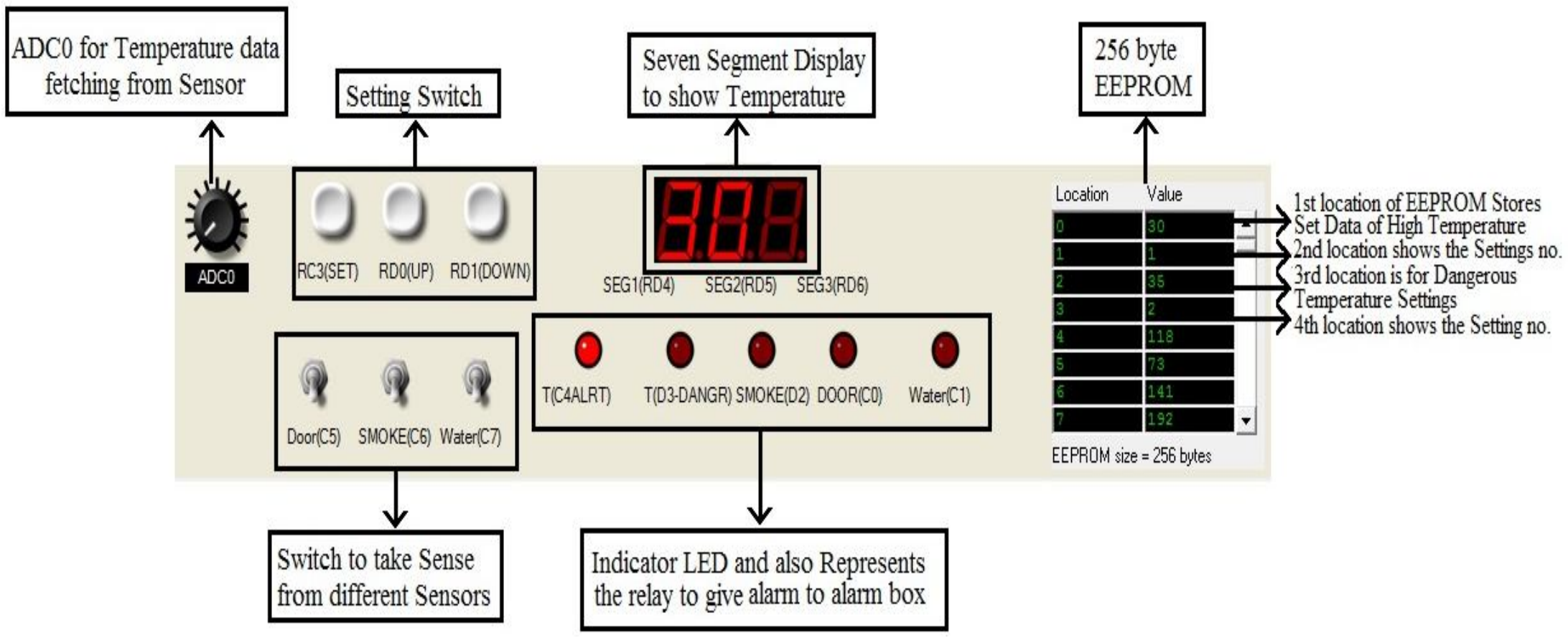

Figure 8. Simulation Pannel

### 4.1 LOGIC FLOW CHART

Initially declare variables for all the sensors and read preset values from EEPROM to those variables. If the set switch is pressed then enter in to the settings micro and change the necessary entities and store those data in to the EEPROM. Now check ADC0 pin for any change in the Temperature and other sensor connected ports as well. If there is any unexpected occurrence identified then controller will send alarm to the relevant port of the alarm box.

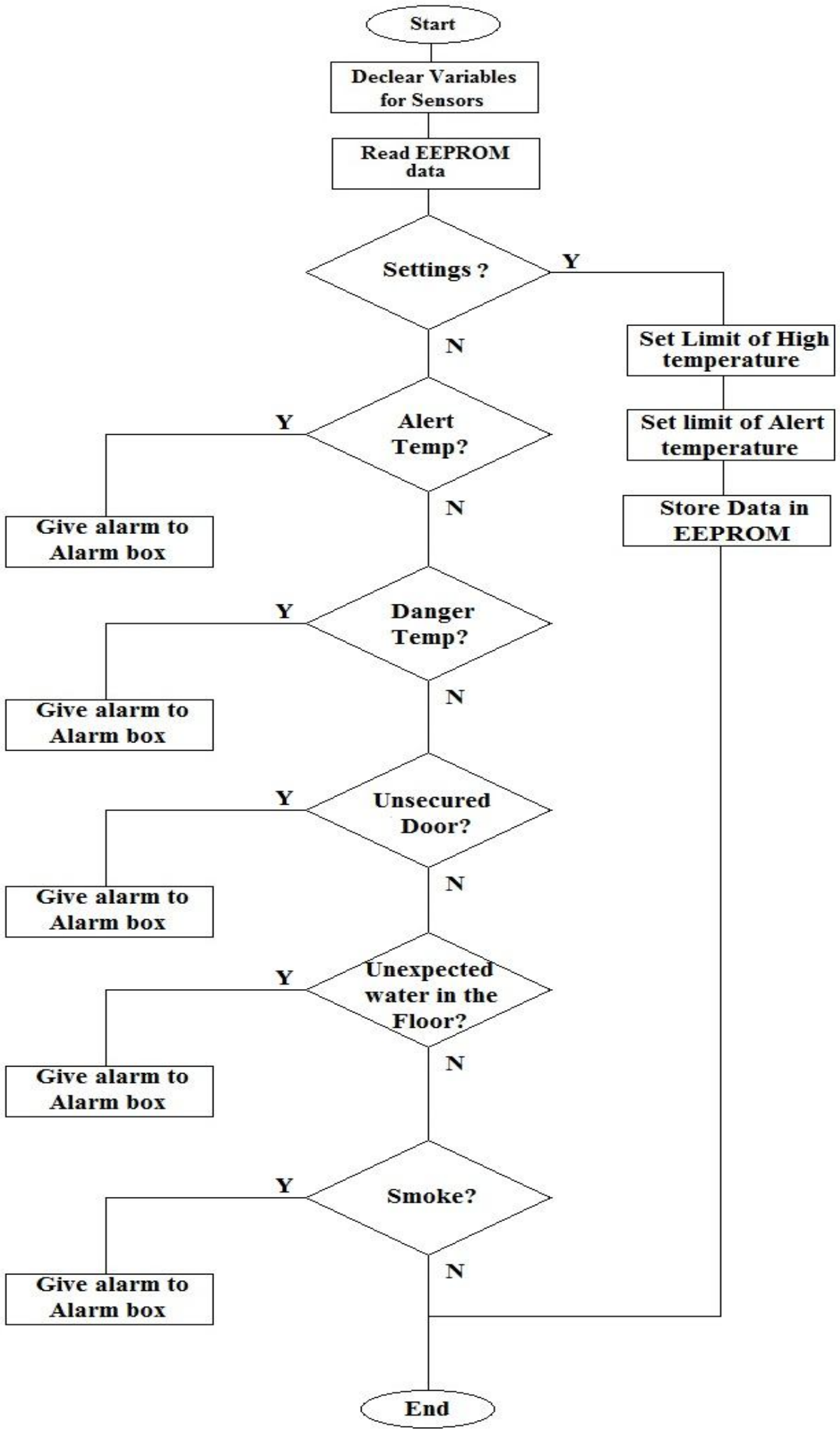

Figure 9. Logical flowchart

## 5. CIRCUIT DIAGRAM

PIC16F877A is used as the main controller where Seven Segment is connected in the PORT B and its common pins are controlled by PORT D. Sensors are connected in RC5, RC6, RC7 and the only ADC is used for temperature sensor and that is connected in AN0. Setting switches are connected in PORT D and Port C. Here some pins of port D and Port C are used to control the relays which are providing the sense to the alarm box. The program is inserted into the PIC by using a software and programmer called USBurn.

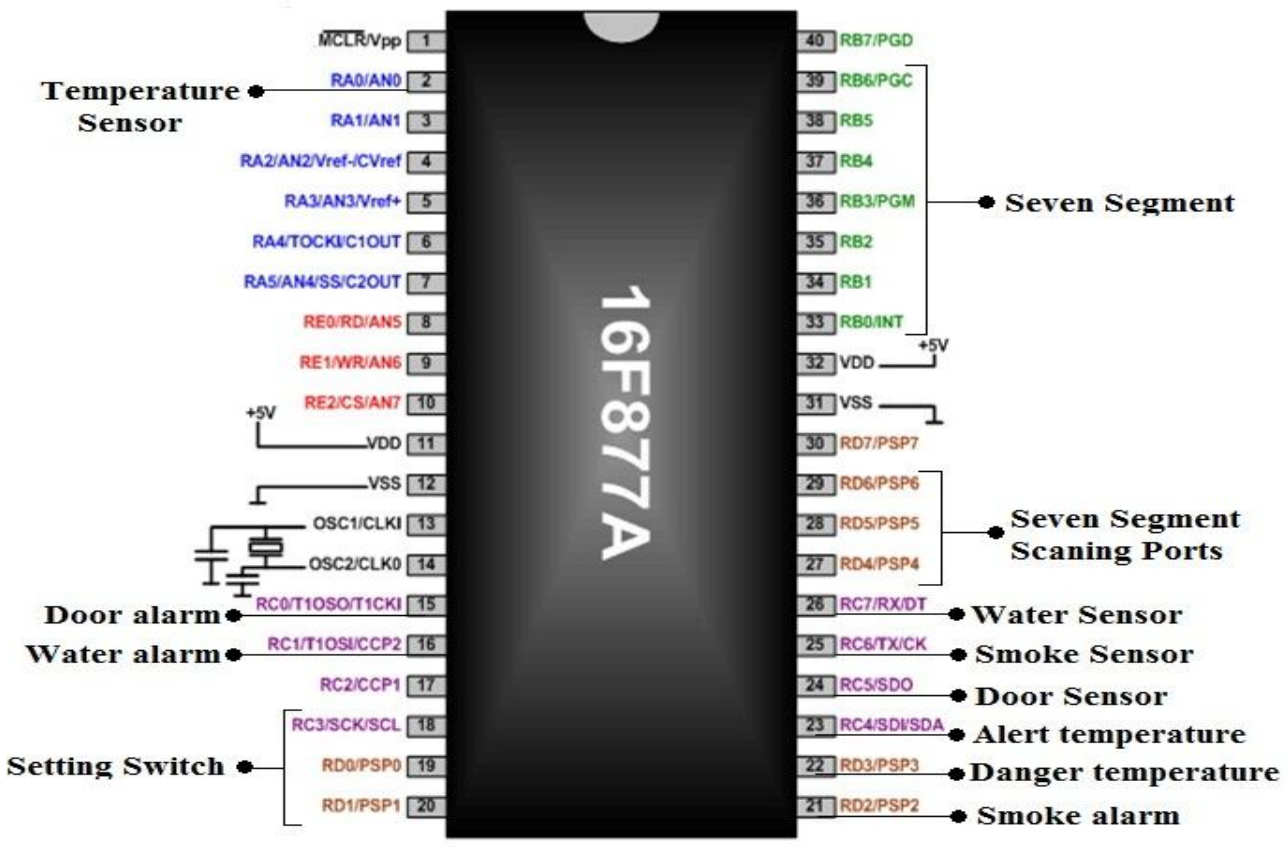

Figure 10. Implemented Circuit Diagram.

## 6. RESULT

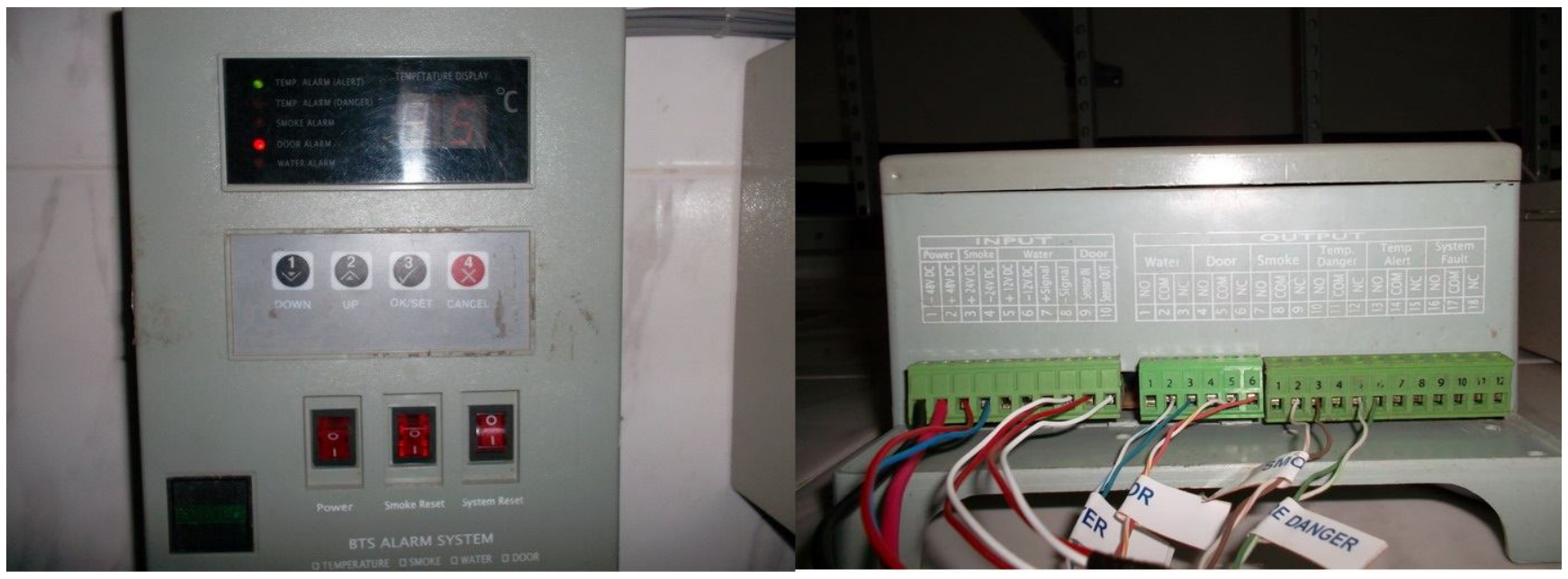

Figure 11. Alarm Controller and Its Connection Ports.

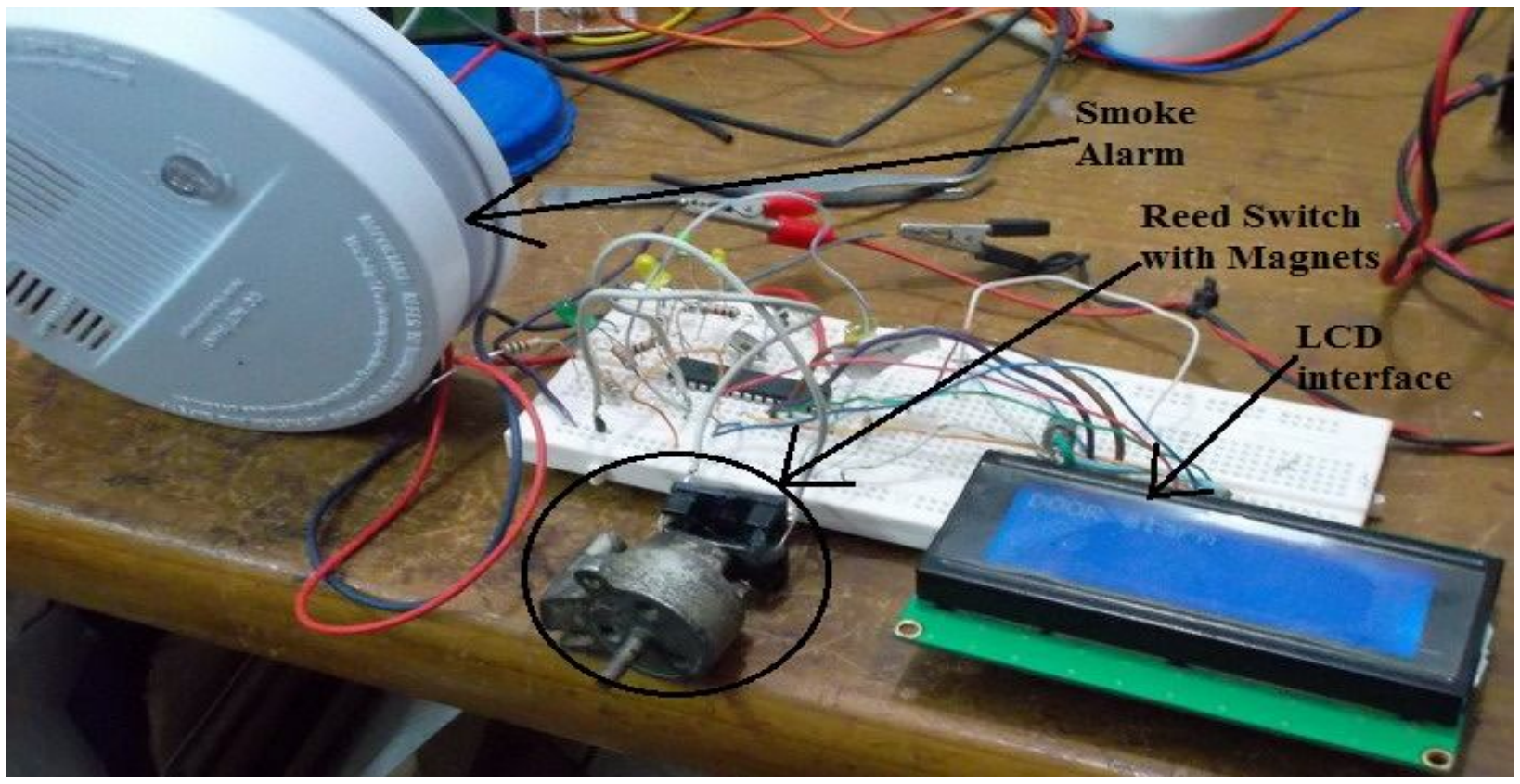

Figure 12: Alarm Circuit testing using LCD and PIC 16F876A

In figure 11 a complete controller is shown which is installed in a BTS site and also its port diagram. In figure 12 a complete alarm circuit is shown where a LCD interface is used and here 16F876A is used for testing different alarms.

## 7. CONCLUSION

Safety of the BTS room is a serious issue over the years as most of the BTS rooms are located in remote places and all of them contain costly and sophisticated equipments. So this designed controller offers a robust safety feature for the BTS room. The 10 bit ADC can give an accurate temperature value which can secure the BTS room from any kind of fire oriented disaster along with the smoke alarm system. For smoke alarm highly sensitive photoelectric sensor based system is used which can give alarm for any kind of sudden smoke sense. For water and door alarm section locally developed sensors are used but their performance is satisfactory and reliable as well. This whole system is low cost so can be affordable as well.

**Authors**

**Arifa Ferdousi**

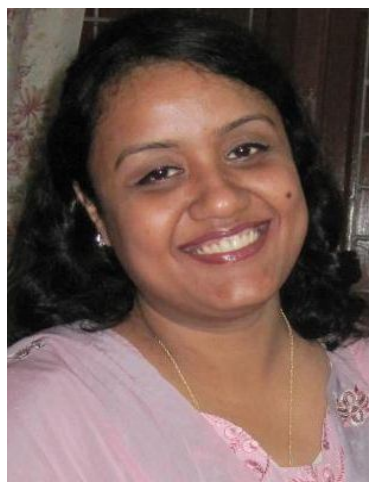

Arifa Ferdousi received B.Sc. and M.Sc. degree in ICE from University of Rajshahi, Bangladesh, in the year of 2007 and 2009 respectively. Currently she is working as a lecturer in the department of CSE in Varendra University, Rajshahi, Bangladesh. Her research interest includes electronics system designing, OFDM, Advanced LTE Wi-Max and Bangla speech recognition system using Neural Network. She is the member of Bangladesh Electronic Society (BES).

**Sadeque Reza Khan**

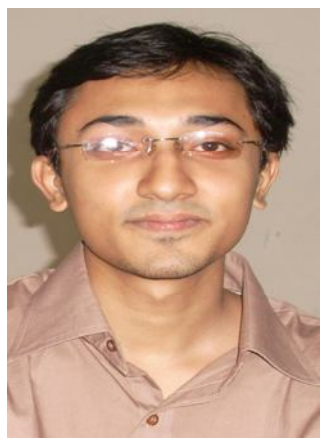

Sadeque Reza Khan received B.Sc. degree in Electronics and Telecommunication Engineering from University of Liberal Arts Bangladesh and continuing his M.Tech in VLSI from National Institute of Technology Karnataka (NITK), India. Currently he is in study leave from his Institution where he was working as a lecturer in the department of Electrical and Electronic Engineering in Prime University, Bangladesh. His research interest includes VLSI, Microelectronics, Control System Designing and Embedded System Designing.